\documentclass[12pt]{article}

 \usepackage{multirow}
 \usepackage{multicol}
\usepackage{amssymb,amsmath, amsfonts}
\usepackage{cases}
\usepackage{latexsym}
\usepackage{graphicx}
\usepackage{relsize}
\usepackage{natbib}
\usepackage{rotating}
\usepackage[includeheadfoot,left=1.in,right=1.in,top=1.in,bottom=1.in,bindingoffset=0.in,nohead]{geometry}
\usepackage[onehalfspacing]{setspace}
\usepackage{booktabs}
\usepackage{appendix}
\usepackage[pdfborder={0 0 0}]{hyperref}
\usepackage{float}
\usepackage{xcolor}
\usepackage{caption}
\usepackage{subcaption}
\usepackage{minibox}

\usepackage[]{listings}
\begin{document}

\title{On the Evolution of  \\ U.S. Temperature Dynamics}

\author{Francis X. Diebold\\University of Pennsylvania \and Glenn D. Rudebusch \\Federal Reserve Bank of San Francisco}

\date{}

\maketitle

\begin{center} 
	\large
	
	This Version, \today

	\normalsize
	
\end{center}

\begin{spacing}{1}

	\bigskip
	
	\large
	\noindent Forthcoming in A. Chudek, C. Hsiao and A Timmermann (eds.), \textit{Essays in Honor of M. Hashem Pesaran} (\textit{Advances in Econometrics}, Volume 44), in press.
	\normalsize
	
	\bigskip

	\bigskip

	\noindent \textbf{Abstract}: Climate change is a massive multidimensional shift. Temperature shifts, in particular, have important implications for urbanization, agriculture, health, productivity, and poverty, among other things. While much research has documented rising mean temperature \emph{levels}, we also examine range-based measures of daily temperature \emph{volatility}. Specifically, using data for select U.S. cities over the past half-century, we compare the evolving time series dynamics of the average temperature level, AVG, and the diurnal temperature range, DTR (the difference between the daily maximum and minimum temperatures). We characterize trend and seasonality in these two series using linear models with time-varying coefficients. These straightforward yet flexible approximations provide evidence of evolving DTR seasonality and stable AVG seasonality.

	\thispagestyle{empty}

	\bigskip

	\noindent {\bf Acknowledgments}:   For comments and/or assistance we are especially grateful to the Editor and the Referee, as well as Bob Amano, Michael Bauer, Sean Campbell, Preston Ching, Philippe Goulet  Coulombe, Rob Engle, Max G\"obel, Jesus Gonzalo, Michael Goldstein, David Hendry, Luke Jackson, Kajal Lahiri, Golnoush Rahimzadeh, Dick Startz, Allan Timmermann, Mike Tubbs,  David Wigglesworth, and Boyuan Zhang, and seminar/conference participants at the Universities of Chicago, Nottingham,  Oxford, and  Washington.

	\bigskip
	
	
	\bigskip

	
	\bigskip
	
	\noindent {\bf Contact}:  fdiebold@upenn.edu, glenn.rudebusch@sf.frb.org

\end{spacing}

%
%
%

\clearpage

\setcounter{page}{1}
\thispagestyle{empty}

\section{Introduction}
Climate change can be defined as the variation in the joint probability distribution describing the state of the atmosphere, oceans, and fresh water including ice \citep{HsiangJEP}. These are complex, multidimensional physical systems, and the various features of climate change have been described using a diverse set of summary statistics. One of the most important aspects of climate change is the evolving distribution of temperature, which has important implications for growth, urbanization, agriculture, health, productivity, and poverty  \citep{Pesaranetal2019, Colacito2019,NBERw27599}.

A diverse set of indicators have been used to measure such distributional temperature variation, including, for example, mean temperature, temperature range, hot and cold spell duration, frost days, growing season length, ice days, heating and cooling degree days, and start of spring dates \citep{IPCC2018, USGCRP2018}. Of course, the \textit{level} of temperature -- the central tendency of the distribution -- has attracted the most attention, in particular regarding the upward trend in the average daily temperature (AVG) \citep{Gonzalo2020}. In contrast, less attention has been given to temperature \textit{volatility}.\footnote{An interesting and novel exception is \cite{Goldstein2017}, based on options-theoretic thinking.}  Temperature volatility can be measured by the diurnal temperature range (DTR), which is the difference between the daily maximum temperature (MAX) and minimum temperature (MIN).

Similar to changes in temperature averages, changes in temperature ranges and variability can also have important effects on environmental and human health \citep{Davyetal2017}.  For example, the incidence of temperature extremes such as heat waves depends critically on how the whole distribution of temperature is shifting -- including both central tendency and variability. Of course, such temperature extremes can have notable adverse effects on society and the economy. Temperature variability can stress workers and lower labor productivity, but it can also have direct effects on output. A salient example is agriculture, whose output is a function of capital, labor, and weather inputs.\footnote{\cite{Wigglesworth2019} finds an important role of DTR in a panel study of U.S. state-level agricultural production over and above standard covariates like capital, labor, and AVG.} Indeed, the very viability of certain agricultural sub-industries, notably wine  and coffee production, is crucially dependent on temperature ranges. For example, \cite{Robinson2006} notes that

\begin{quotation}

 \noindent Diurnal temperature variation is of particular importance in viticulture. Wine regions situated in areas of high altitude experience the most dramatic swing in temperature variation during the course of a day. In grapes, this variation has the effect of producing high acid and high sugar content as the grapes' exposure to sunlight increases the ripening qualities while the sudden drop in temperature at night preserves the balance of natural acids in the grape. \quad  (p. 691)

 \end{quotation}

To better understand the full nature of the changing distribution of temperature, we examine DTR in select cities in the United States over the past half-century.  We allow for time-varying coefficients, which provide a straightforward yet flexible approximation to more general nonlinear effects. Although our focus is on DTR, we also provide a parallel analysis for AVG, which allows valuable interpretive context and contrast. Our work reveals an \textit{evolving} DTR conditional mean seasonal pattern, in contrast to the fixed AVG conditional mean seasonal pattern. 

The previous research literature that examined DTR struggled for some time to develop firm conclusions about the dynamics of temperature variability. Even the direction of the trend in DTR has been somewhat contentious \citep{Alexander2013}. Recent work has established that the downward trend in DTR in many locations \citep{Sunetal2019} reflects a more rapid warming of MIN than MAX -- generally the result of nighttime lows rising faster than daytime highs \citep{Davyetal2017}. However, this differential trending of MIN and MAX, or ``diurnal asymmetry,'' is not geographically uniform because of variation in vegetation, cloud cover, and other factors \citep{Jacksonetal2010, Sun2014}. Along with this trend in temperature variability, seasonal variation in DTR has also been considered by a few authors who describe a lower temperature range in winter than at other times  \citep{Ruschyetal1991,Durreetal2001}. There is  also some evidence that the seasonality of DTR in the United States may be changing over time \citep{Quetal2014}. To capture as much variation as possible in the distribution of DTR -- including trend and seasonal -- we use linear time series models with time-varying coefficients to provide simple yet powerful representations.

We proceed as follows. In section \ref{basic}, we provide an introductory analysis for a representative city, Philadelphia.  Then, in section \ref{fifteen}, we  broaden the analysis to include fifteen geographically dispersed U.S. cities. We conclude in section \ref{concl}.

\section{Philadelphia}  \label{basic}

We introduce and illustrate our approach by studying temperature data measured at the Philadelphia airport (PHL) in a step-by-step fashion.  We present most results graphically, while regression results on which these graphs are based appear in Appendix \ref{regresphl}.\footnote{EViews code is available at \url{https://www.sas.upenn.edu/~fdiebold/papers/paper122/DTRcode.txt}} The underlying data are the daily MAX and MIN measured in degrees Fahrenheit, obtained from the U.S. National Ocean and Atmospheric Administration's Global Historical Climate Network database (GHCN-daily).\footnote{The data are available at \url{https://www.ncdc.noaa.gov/ghcn-daily-description}. For details, see \cite{Menneetal2012} and \cite{Jaffres2019}.} Our sample period is from 01/01/1960 to 12/31/2017, which covers the period of almost all recent climate change.

\subsection{Distributions}

\begin{figure}[tb]
	\caption{Estimated Densities, AVG and DTR, Philadelphia}
	\begin{center}
		\includegraphics[trim= 6mm 0mm 0mm 0mm, clip, scale=.5]{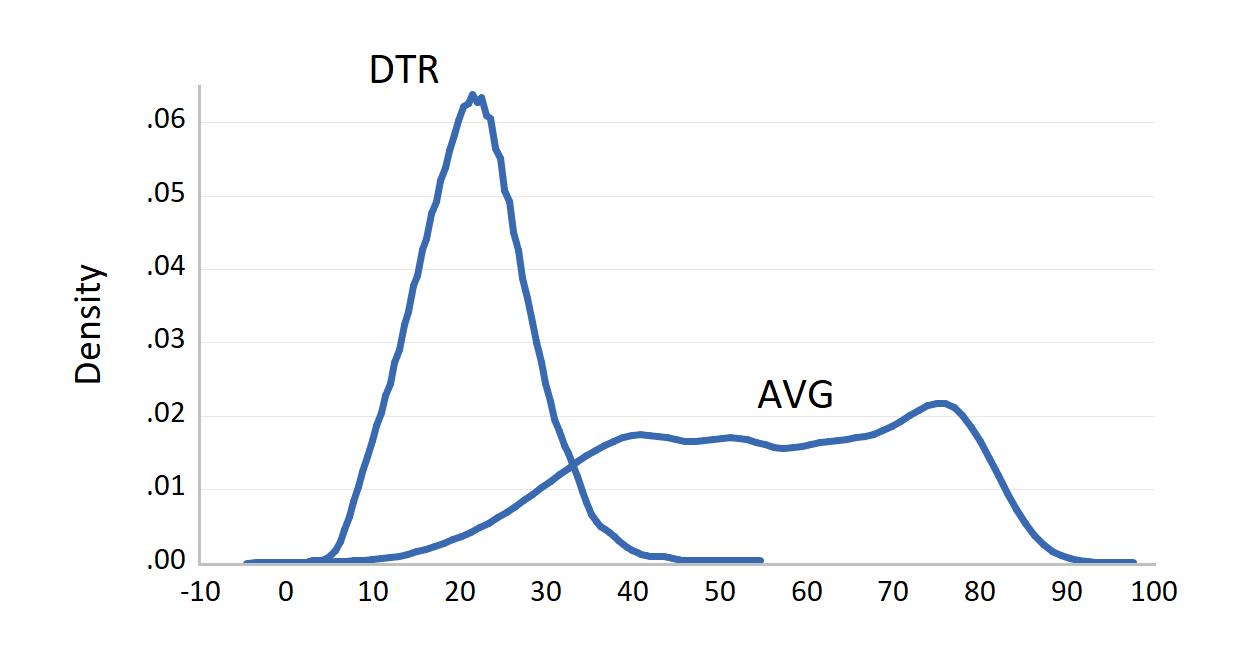}
		\label{dens}
	\end{center}
		\footnotesize{Notes to figure: We show kernel  density estimates for daily AVG and DTR, 1960-2017.}
\end{figure}

The daily MAX and MIN are informative of both the central tendency and variability of the daily continuous-time temperature record. In particular, the daily average temperature, AVG=(MAX+MIN)/2, is a natural measure of central tendency, and the daily temperature range, DTR=MAX-MIN, is a natural measure of volatility or variability.\footnote{AVG and DTR are standard measures, used almost universally.  Of course one could entertain more sophisticated measures of central tendency than AVG, for example, but again AVG and DTR are standard.}  DTR is a natural and intuitive estimator of daily volatility, and it is also highly efficient statistically. The ``daily range'' has a long and distinguished tradition of use in econometrics due to its good properties in estimating underlying quadratic variation from discretely-sampled data \citep{ABD2002}. Interestingly, although AVG has been studied extensively \citep{Rafteryetal2017}, DTR has been studied much less.

In Figure \ref{dens}, we show kernel  estimates of the unconditional densities of AVG and DTR. The bimodal shape of the AVG density reflects the strong seasonality in AVG. The ``winter mode" is around 40$^{\circ}$F, and the ``summer mode" is around 75$^{\circ}$F.  The AVG density contrasts sharply with the unimodal approximately-symmetric density of DTR, which is centered around 19$^{\circ}$F and much less dispersed.

\subsection{Trend}

\begin{figure}[tb]
	\caption{Data and Estimated Trends, AVG and DTR,  Philadelphia}
	\begin{center}
		\includegraphics[trim= 6mm 0mm 0mm 0mm, clip, scale=.9]{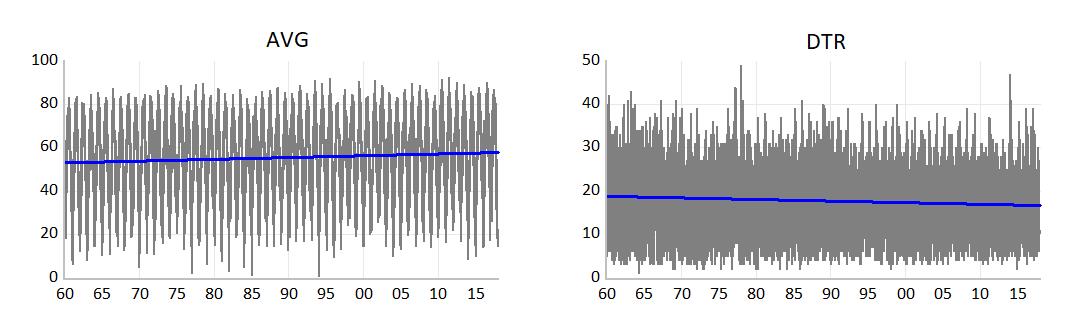}
		\label{tsplots}
	\end{center}
		\footnotesize{Notes to figure: We show time series of daily AVG and DTR (gray) together with estimated linear trends (blue), 1960-2017.  The vertical  axes are scaled differently in the two panels, and they are in degrees Fahrenheit.}
\end{figure}

In Figure \ref{tsplots}, we display time series plots of the entire data sample of AVG and DTR with fitted linear trends superimposed. The regression is
\begin{equation} \label{trendline}
Y \rightarrow c, TIME,
\end{equation}
where $Y$ is AVG or DTR, $c$ is a constant, and $TIME$ is a time trend (that is, $TIME_t = t$ and $t=1, ..., T$).  Here and throughout, we use  heteroskedasticity and autocorrelation consistent (HAC) standard errors to assess statistical significance \citep{nw1987}.

The AVG trend slopes upward and is statistically significant, which is consistent with the overall global warming during this period. The steepness of this trend is surprising, as the AVG trend grows by nearly five degrees Fahrenheit over the course of the 57-year 1960-2017 sample.  This increment is a bit more than twice as much as the average global increase over the same period \citep{rudebusch2019EL}. The faster upward trend in the Philadelphia airport average temperature likely reflects two key factors: (1) average temperatures in growing cities tend to rise more quickly due to an increasing urban heat island effect and (2) average land temperatures generally grow more quickly than the global average, which includes ocean areas that are slow to warm.

As for Philadelphia temperature variability, DTR also has a significant trend, and it slopes  \textit{downward}, dropping  by more than two degrees over the course of the sample. The downward DTR trend arises from different trends in  the underlying MAX and MIN  -- a diurnal asymmetry.  Both trend upward, but MIN is on a steeper incline as evening temperatures warm more quickly. Hence, the spread between MAX and MIN tends to shrink, and DTR decreases over time.  The relatively muted upward trend in MAX is generally ascribed to increased cloud cover, soil moisture, and precipitation, which lead to diminished incoming solar radiation and increased daytime surface evaporative cooling; however, with local variation in these meterological elements, a downward DTR trend is not found at all locations \citep{Dai1999, Davyetal2017,Vinnarasietal2017}.

The overall picture, then, involves not only an upward trend in AVG, but also a gradual tightening of daily fluctuations around that trend. Warming is not only happening, but also happening more reliably.


\subsection{Fixed Seasonality}  \label{fixseas}

\begin{figure}[tb]
	\caption{De-Trended Data and Estimated Fixed Seasonals, AVG and DTR, Philadelphia}
	\begin{center}
		\includegraphics[trim= 6mm 0mm 0mm 0mm, clip, scale=.9]{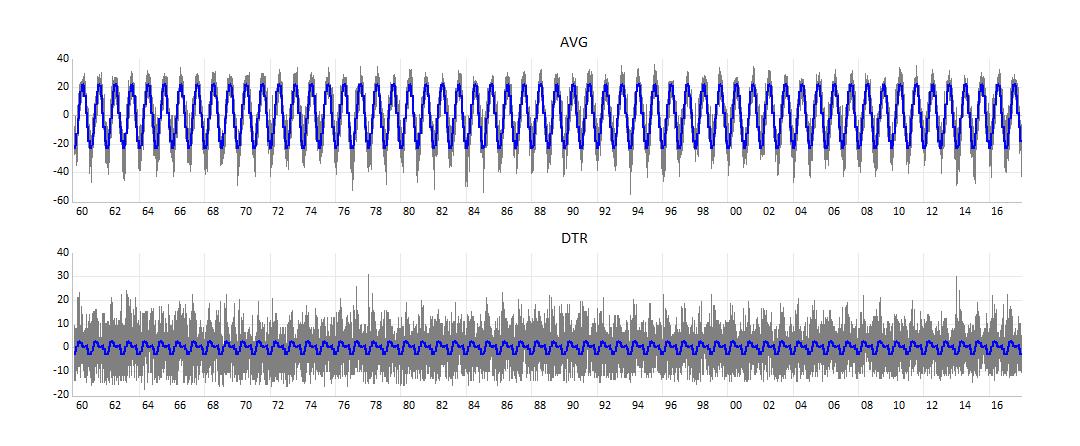}
		\label{plotfit}
	\end{center}
	\footnotesize{Notes to figure: We show time series of daily linearly de-trended AVG and DTR (gray) together with  estimated fixed seasonals (blue) from regressions of daily linearly de-trended data on 12 monthly seasonal dummies, 1960-2017.   The vertical and horizontal axes are scaled identically in the top and bottom panels. The vertical  axes are in degrees Fahrenheit.}
\end{figure}

 In Figure \ref{plotfit}, we show the actual and fitted values from regressions of de-trended AVG and DTR on 12 monthly seasonal dummies,
 \begin{equation}
 \label{seasreg}
\widetilde{Y} ~ \rightarrow~ D_1, ..., D_{12},
\end{equation}
where $\widetilde{Y}$ is de-trended AVG or DTR -- the residuals from regression (\ref{trendline}) -- and $D_{it}=1$ if day $t$ is in month $i$, and 0 otherwise.\footnote{There is of course no need for an  intercept, which would be completely redundant and hence cause perfect multicollinearity.}  This model is effectively an intercept regression for deviations from trend, allowing for a different intercept each month.

As shown in the top panel of Figure \ref{plotfit}, AVG displays pronounced seasonality. The seasonality is highly significant and is  responsible for a large amount AVG variation. The $R^2$ of the seasonal  AVG regression (\ref{seasreg}) is .81. As with the upward trend in AVG, strong seasonality in deviations of AVG from its trend is hardly surprising -- it's cold in the winter and hot in the summer.

There is also significant seasonality in DTR, as shown in the bottom panel of Figure \ref{plotfit}. The    DTR seasonality was  hard to detect visually in the time series plot of Figure \ref{tsplots}, because it is  buried in much more noise than that of  AVG. The $R^2$ of the seasonal DTR   regression  (\ref{seasreg}) is only .07.

\begin{figure}[tb]
	\caption{Estimated Fixed Twelve-Month Seasonal Patterns, AVG and DTR,  Philadelphia}
	\begin{center}
		\includegraphics[trim= 6mm 8mm 0mm 0mm, clip,scale=.25]{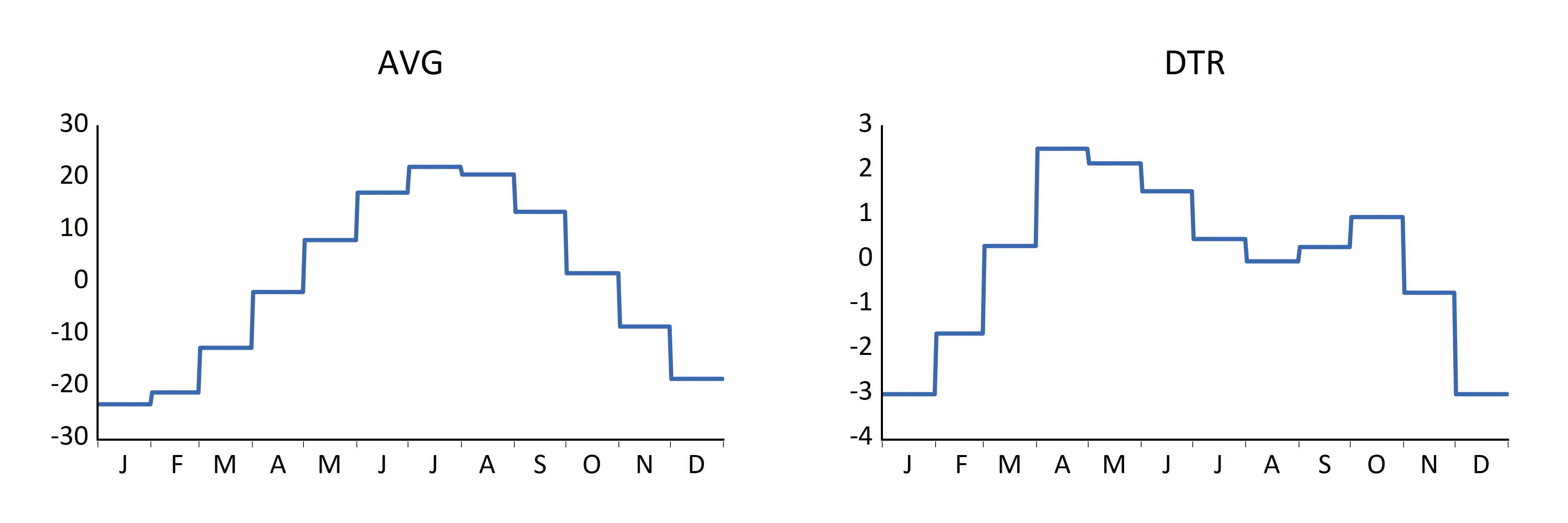}
		\label{seas}
	\end{center}
		\footnotesize{Notes to figure: We show  estimated fixed twelve-month seasonal patterns for AVG and DTR, based on regressions of daily linearly de-trended data on 12 monthly seasonal dummies, 1960-2017. 	The vertical  axes are scaled differently in the left and right panels, and they are in degrees Fahrenheit.}
\end{figure}

In Figure \ref{seas}, we show the estimated monthly seasonal factors for  AVG (left panel) and DTR (right panel).  They are simply the  12 estimated coefficients on the 12 monthly dummies in the seasonal regression (\ref{seasreg}).  The seasonal pattern for AVG is as expected -- smooth and unimodal, high in the summer and low in the winter, achieving its maximum in July and its minimum in January. In contrast, the  seasonal pattern for DTR is clearly bi-modal, with one mode in April-May and one in October. DTR's two annual peaks (spring and fall) and two annual troughs (winter and summer)  contrast sharply with AVG's single annual peak (summer) and single annual trough (winter). This ``twin-peaks" or ``M-shaped" DTR pattern is common across many U.S. cites. Moreover, as we shall show, in many locations, the DTR seasonal pattern has evolved noticeably over time with climate change, whereas the AVG seasonal pattern has remained stable.

\subsection{Evolving Seasonality}

\begin{figure}[tb]
	\caption{Estimated Evolving Twelve-Month Seasonal Patterns, DTR and AVG, Philadelphia, 1960 vs. 2017}
	\begin{center}
		\includegraphics[trim= 6mm 8mm 0mm 0mm, clip, scale=.25]{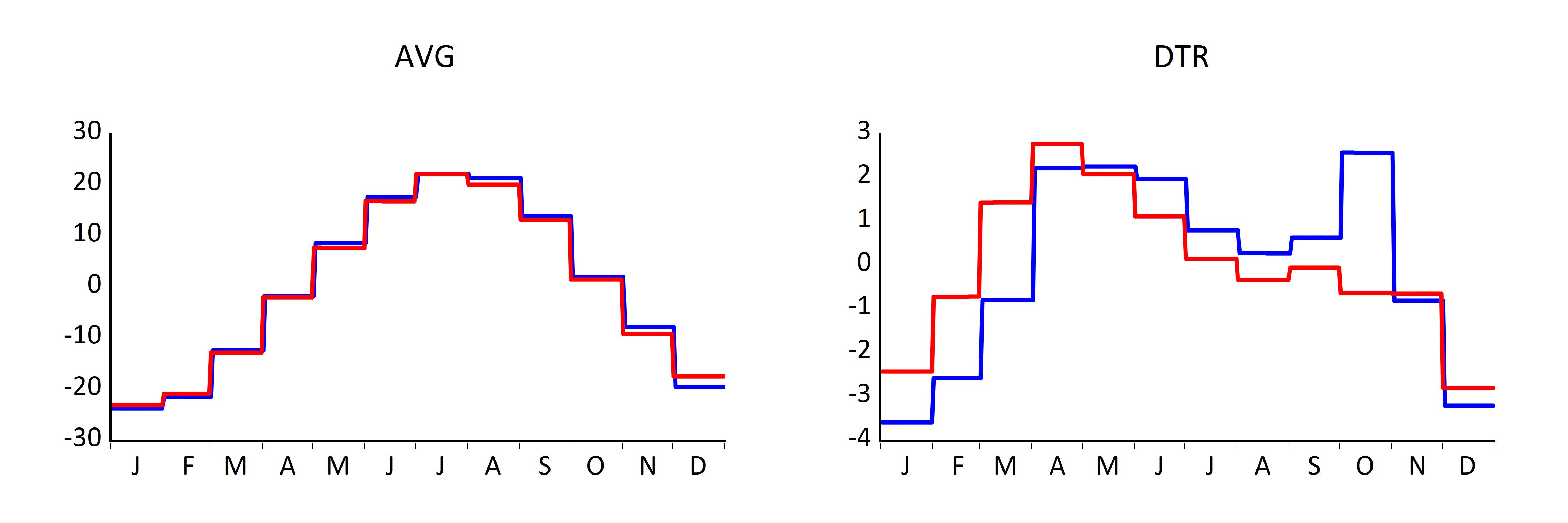}
		\label{trendfit222b}
	\end{center}
	\footnotesize{Notes to figure: We show the estimated twelve-month   seasonal patterns  of  AVG and DTR,  based on  regressions of daily linearly de-trended data on 12 monthly seasonal dummies,  and those same dummies interacted with time, 1960-2017.   1960 is blue, and 2017 is red. The vertical  axes are scaled differently in the left and right panels, and they are in degrees Fahrenheit.}
\end{figure}

The AVG and DTR trends documented thus far are trends in  \textit{level}. More subtle are trends in \textit{seasonality} -- that is, trends in the tent-shaped AVG seasonal pattern and the M-shaped DTR seasonal pattern. We now explore the possibility of such evolving seasonality by allowing for linear trends in the seasonal factors.\footnote{Our framework of linear trend in  seasonal factors  is quite flexible, and in any event much more flexible than what is routinely entertained in the climatological literature.  One could allow even more flexibility by allowing, for example,  possible quadratic trend in seasonal factors, as done in a different context by \cite{DRice}.}  Mechanically, this involves regressing de-trended AVG or DTR not only on 12 monthly dummies, but also those same 12 dummies interacted with time,
\begin{equation}  \label{evolvereg}
\widetilde{Y} ~\rightarrow ~ D_1, ..., D_{12}, ~~ D_1 {\cdot} TIME, ..., D_{12} {\cdot} TIME,
\end{equation}
where $\widetilde{Y}$ is de-trended AVG or DTR, $D_{it}=1$ if day $t$ is in month $i$ and 0 otherwise, and $TIME_t = t$. Regression  (\ref{evolvereg}) can capture linearly-trending seasonal deviations from a linear trend. Effectively, it  allows for a different intercept each month, with those intercepts themselves potentially trending at different rates. In the special case where all interaction coefficients are zero, it collapses to fixed seasonal deviations from  linear trend, as explored in section \ref{fixseas}.

For AVG, there are no gains from estimating the more flexible seasonal specification (\ref{evolvereg}). The interaction terms are universally insignificantly different from zero, clearly indicating no change over time in the AVG seasonal pattern.  In the left panel of Figure \ref{trendfit222b}, we show the estimated seasonal factors for AVG for the  first year (1960) and last year (2017) of our sample.  This range provides the maximum contrast, but the two seasonal patterns are nevertheless essentially identical.

The results for DTR, however, are very different. Unlike the AVG seasonal, which does not evolve, the DTR seasonal changes significantly over time. The January-through-March DTR interaction coefficients are significantly positive, indicating that the winter DTR low is increasing.  In addition, all May-through-October interaction coefficients are negative, and the October coefficient is large and highly significantly negative. This corresponds to progressively lower DTR highs in Octobers, so that the fall  DTR peak  is gradually vanishing. Both effects (higher winter DTR lows, and lower fall DTR highs) are visually apparent in the right panel of Figure \ref{trendfit222b}, in which we contrast the estimated DTR M-shaped seasonal pattern in the first year (1960) and last year (2017) of our sample.

\section{Fifteen Cities} \label{fifteen}

\begin{figure}[tb]
	\caption{Fifteen Cities}
	\begin{center}
		\includegraphics[trim= 0mm 0mm 0 0mm, clip, scale=.70]{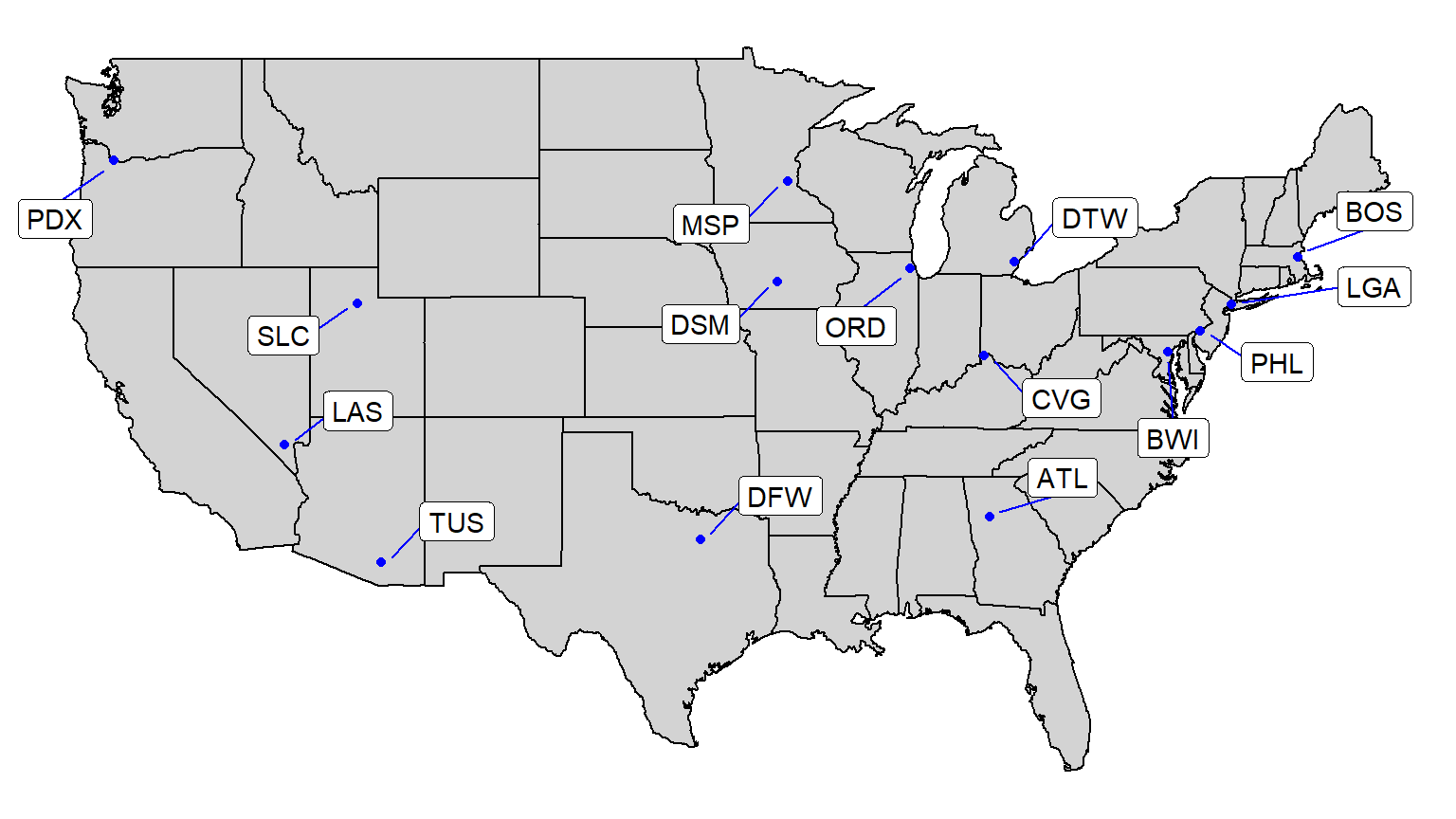}
		\label{cities}
	\end{center}
	\footnotesize{Note to figure:  We show the fifteen cities for which we study AVG and DTR, by airport code.}
\end{figure}

We now expand our analysis to include data from the airports of the fifteen U.S. cities shown in Figure \ref{cities}.\footnote{We emphasize that we study only urban areas. They are of maximal  interest because a large fraction of the population is concentrated there.  However, urban temperature patterns and their evolution (e.g., for AVG or DTR) may (and do) of course differ from those in non-urban areas.}   As with the Philadelphia case study in section \ref{basic}, we obtain the underlying daily MAX and MIN data, from which we construct daily AVG and DTR, from the U.S. National Ocean and Atmospheric Administration's GHCN-daily, \url{https://www.ncdc.noaa.gov/ghcn-daily-description}.  Our sample period is 01/01/1960-12/31/2017.\footnote{There were a (very) few missing observations, in which case we interpolated using an average of the immediately previous and subsequent days' values, rounded to the nearest integer. The missing observations are: BWI max: 1/7/04, min: 1/6/04, DSM max: 9/15/96, min: 9/15/96, and TUS max: 5/10/10, 8/18/17, 8/19/17, min: 5/11/10, 8/18/17, 8/19/17.}

We choose these city weather reporting stations because all of them have had temperature derivatives traded  on the Chicago Merchantile Exchange (CME) \citep{CD2005}. Consideration of such CME cities is of interest for several reasons.  First, these locations cover a diverse set of climates, so they can provide a check of the robustness of our Philadelphia results. Second, they are urban locations that represent large numbers of people and a sizable share of economic activity -- one reason that their CME contracts are traded.  Finally, the valuations of weather derivatives traded in financial markets depend on the evolution of the stochastic structure of temperature dynamics, which is precisely the focus of our modeling efforts and so naturally paired with the CME cities.

The full set of historically-traded cities includes: Atlanta, ATL; Boston, BOS; Baltimore Washington, BWI; Chicago, ORD; Cincinnati, CVG; Dallas Fort Worth, DFW; Des Moines, DSM;  Detroit, DTW;  Houston, IAH; Kansas City, MCI; Las Vegas, LAS; Minneapolis St Paul, MSP; New York, LGA; Portland, PDX; Philadelphia, PHL; Sacramento, SAC;  Salt Lake City, SLC, and Tuscon, TUS. We exclude Houston, Kansas City, and Sacramento, however, due to  large amounts of missing data, leaving fifteen cities. Presently eight cities are traded (Atlanta, Chicago, Cincinnati, Dallas, Las Vegas, Minneapolis, New York, and Sacramento), and all but Sacramento are in our fifteen.\footnote{See \url{https://www.cmegroup.com/trading/weather/temperature-based-indexes.html}.}

We focus on urban airports not only because they are focal points for the financial weather derivatives community, but also because  they are focal points for the  meteorological /  climatological measurement community.  The U.S. National Weather Service surface weather observation network, in particular, is located largely at airports. On balance, airports, with large open spaces of grassland without surrounding buildings, are regarded as fairly representative locations for measuring temperature.


\begin{table} [t]
		\begin{center}
		\newcommand\T{\rule{0pt}{2.6ex}}
		\caption{AVG,  Fifteen Cities}
		\label{vvv}
		\begin{tabular}{c c  c c c c  c c }
			\toprule
			(1) & (2)  & (3) &  (4)  & (5)  & (6)   & (7)    \\
			$station$ & $\Delta trend$  & $p(nt)$  &  $p(ns)$ &  $p(nts)$ & $\rho$  & {$R^2$}    \\
			\midrule
			ATL     & 4.36$^*$ & 0.00     & 0.00     & 0.99      & 0.76$^*$ & 0.90 \\
			BOS     & 2.06$^*$      & 0.00     & 0.00     & 0.73   & 0.67$^*$ & 0.89 \\
			BWI     & 2.25$^*$      & 0.00     & 0.00     & 0.80   & 0.71$^*$ & 0.90 \\
			CVG     & 2.53  & 0.04  & 0.00     & 0.94   & 0.74$^*$ & 0.89 \\
			DFW     & 3.44$^*$  & 0.00     & 0.00     & 0.55   & 0.72$^*$ & 0.89 \\
			DSM     & 3.93$^*$      & 0.00     & 0.00     & 0.17   & 0.76$^*$ & 0.91 \\
			DTW     & 4.09$^*$      & 0.00     & 0.00     & 0.99   & 0.74$^*$ & 0.91 \\
			LAS     & 6.05$^*$  & 0.00     & 0.00     & 0.41   & 0.82$^*$ & 0.96 \\
			LGA     & 4.03$^*$      & 0.00     & 0.00     & 0.97   & 0.71$^*$ & 0.91 \\
			MSP     & 4.72$^*$      & 0.00     & 0.00     & 0.18   & 0.77$^*$ & 0.93 \\
			ORD     & 2.86$^*$      & 0.00     & 0.00     & 0.78   & 0.74$^*$ & 0.90 \\
			PDX     & 2.55$^*$  & 0.00     & 0.00     & 0.26   & 0.76$^*$ & 0.90 \\
			PHL     & 4.78$^*$  & 0.00     & 0.00     & 0.95   & 0.72$^*$ & 0.91 \\
			SLC     & 3.92$^*$      & 0.00     & 0.00     & 0.67   & 0.77$^*$ & 0.93 \\
			TUS     & 4.89$^*$  & 0.00     & 0.00     & 0.33   & 0.79$^*$ & 0.93 \\
			\midrule
			Median & 3.93 & 0.00 &	0.00 &	0.73 &	0.74 &	0.91 \\
			\bottomrule
		\end{tabular}
	\end{center}
\footnotesize 
\begin{singlespace}
		\noindent  	Notes to table: All results are based on daily data, 1960-2017.  Column 1 reports measurement station by airport code. Column 2 reports the  estimated  trend movement over the entire 57-year sample in degrees Fahrenheit, using a simple regression on linear trend. 	The remaining columns  report results from the  conditional-mean regression (\ref{evolvereg2}).  $p(nt)$ is the robust $p$-value for  a Wald test of no trend (all coefficients on TIME and  $D {\cdot} TIME$ interactions are 0), $p(ns)$ is the robust $p$-value for a  Wald test of no seasonality  (all coefficients on $D$'s and  $D {\cdot} TIME$ interactions are 0),  and $p(nts)$ is the robust $p$-value for  Wald  a test of no trend in seasonality  (all coefficients on  $D {\cdot} TIME$ interactions are 0). $\rho$ is the estimated autoregressive coefficient, and $R^2$ is the coefficient of determination.  Asterisks denote significance at the one percent level.  See text for details.
	\end{singlespace}
\normalsize
\end{table}

\begin{table} [tb]
	\begin{center}
		\newcommand\T{\rule{0pt}{2.6ex}}
		\caption{DTR,  Fifteen Cities}
		\label{vvv2}
		\begin{tabular}{c c  c c c c  c c }
			\toprule
			(1) & (2)  & (3) &  (4)  & (5)  & (6)   & (7)    \\
			$station$ & $\Delta trend$  & $p(nt)$  &  $p(ns)$ &  $p(nts)$ & $\rho$  & {$R^2$}    \\
			\midrule
			ATL     & -1.65$^*$  & 0.00     & 0.00     & 0.14   & 0.38$^*$ & 0.18 \\
			BOS     & -0.48$^*$      & 0.00     & 0.00     & 0.00      & 0.25$^*$ & 0.10 \\
			BWI     & -0.43      & 0.34  & 0.00     & 0.50   & 0.38$^*$ & 0.19 \\
			CVG     & -1.31$^*$  & 0.00     & 0.00     & 0.04   & 0.32$^*$ & 0.17 \\
			DFW     & -1.31$^*$  & 0.00     & 0.00     & 0.64   & 0.40$^*$ & 0.17 \\
			DSM     & -0.51$^*$      & 0.00     & 0.00     & 0.03   & 0.32$^*$ & 0.15 \\
			DTW     & -2.88$^*$  & 0.00     & 0.00     & 0.00      & 0.33$^*$ & 0.27 \\
			LAS     & -7.02$^*$  & 0.00     & 0.00     & 0.13   & 0.46$^*$ & 0.37 \\
			LGA     & 0.03$^*$       & 0.00     & 0.00     & 0.00      & 0.23$^*$ & 0.14 \\
			MSP     & -3.07$^*$  & 0.00     & 0.00     & 0.00      & 0.31$^*$ & 0.18 \\
			ORD     & -2.03$^*$  & 0.00     & 0.00     & 0.00      & 0.30$^*$ & 0.20 \\
			PDX     & -1.68$^*$  & 0.00     & 0.00     & 0.63   & 0.50$^*$ & 0.45 \\
			PHL     & -2.13$^*$  & 0.00     & 0.00     & 0.00      & 0.34$^*$ & 0.19 \\
			SLC     & -4.21$^*$  & 0.00     & 0.00     & 0.00      & 0.44$^*$ & 0.47 \\
			TUS     & 0.48       & 0.05  & 0.00     & 0.03   & 0.51$^*$ & 0.35 \\
			\midrule
			Median & -1.65 &	0.00 &	0.00 &	0.03 &	0.34 &	0.19  \\
			\bottomrule
		\end{tabular}
	\end{center}
	\noindent  \footnotesize
	Notes to table: See Table \ref{vvv}.
	\normalsize
\end{table}

We view the sequential modeling approach employed in section \ref{basic} -- fitting a trend and then characterizing seasonality in the de-trended data -- as intuitive and transparent.  We now consolidate and extend various aspects of that approach, to arrive at a simple yet powerful joint model.
Regarding consolidation, we move from a multi-step sequential modeling approach to a single-step joint approach with a single conditional mean estimation. Regarding extension, we now include an autoregressive lag in the model, which facilitates simple assessment of the strength of serial correlation in the deviations from the trend/seasonal. The autoregressive lag also provides valuable  pre-whitening for serial correlation in HAC covariance matrix estimation \citep{Andrews1992}.

We proceed by regressing AVG or DTR on an intercept, a linear trend, a first-order autoregressive lag, 11 monthly seasonal dummies to capture seasonal intercept variation (we drop July, so the included constant captures July and all estimated seasonal effects are relative to July), and 11 trend-seasonal interactions to capture seasonal trend slope variation (we drop the July interaction):
\begin{equation}  \label{evolvereg2}
\footnotesize Y ~\rightarrow ~ c, TIME, Y(-1), D_1, ..., D_6, D_8, ..., D_{12}, D_1 {\cdot} TIME, ..., D_{6} {\cdot} TIME, D_{8} {\cdot} TIME, ..., D_{12} {\cdot} TIME,
\end{equation}
where $Y$ is AVG or DTR,  $TIME_t = t$,  $Y(-1)$  denotes a 1-day lag, and $D_{it}=1$ if day $t$ is in month $i$ and 0 otherwise.  The joint model (\ref{evolvereg2}) allows for different intercepts each month, with the different  intercepts  potentially trending linearly at different rates, and for serially correlated deviations from the trend/seasonal. We summarize the estimation results in Tables \ref{vvv} and \ref{vvv2}, in which we  show the weather station identifier (airport code) in column 1, and various aspects of the estimation results in subsequent columns.\footnote{Detailed regression results for all cities are in the online Appendix \ref{online} (\url{https://www.sas.upenn.edu/~fdiebold/papers/paper122/OnlineAppendix.pdf},  and underlying EViews code is at \url{https://www.sas.upenn.edu/~fdiebold/papers/paper122/DTRcode.txt}.}


	\subsection{Trend}
	
As shown in column 2 of Table \ref{vvv}, the estimated AVG trend movements over the full sample are large and positive in each city. They are also all highly statistically significant (column 3), with a median $p$-value  of 0.00 for Wald tests of the null hypothesis of no trend. These  $p$-values are denoted $p(nt)$, where ``$nt$" stands for ``no trend", which corresponds to zero coefficients on TIME and all TIME interactions in regression (\ref{evolvereg2}) (in which case it collapses to seasonal intercepts with serial correlation). The median estimated trend movement is 3.38$^{\circ}$F, greater than the consensus estimate of the increase in the mean global temperature over the same period, as U.S. airports have warmed more quickly than the global average.
				
		Similarly, in column 2 of Table \ref{vvv2}, we report the estimated full-sample trend movements  for DTR.  All but one are negative, and most are significant at the one percent level.   The median estimated trend movement is -1.45$^{\circ}$F, with a median $p$-value, $p(nt)$, of 0.00 for the no-trend null hypothesis (column 3). Interestingly, LAS, which has the steepest \textit{upward} AVG trend, also has the steepest \textit{downward} DTR trend.

\subsection{Seasonality}

In column 4 of Tables \ref{vvv} and \ref{vvv2}, we report $p$-values for Wald tests of the hypothesis of no  AVG and DTR seasonality, respectively. These  $p$-values are denoted $p(ns)$, where ``$ns$" stands for ``no seasonality", which corresponds to zero coefficients on all included seasonal dummies and dummy interactions in regression (\ref{evolvereg2}) (in which case it collapses to linear trend with serial correlation). There is of course strong evidence of seasonality in AVG with all $p(ns)$'s equal to 0.00. Less well known is the similarly strong seasonality in DTR with all $p(ns)$'s again equal to 0.00.
		
In column 5 of Tables \ref{vvv} and \ref{vvv2}, we report $p$-values for Wald tests of the hypothesis of no evolving (i.e., trending) AVG and DTR seasonality, respectively. These $p$-values are denoted $p(nts)$, where ``$nts$" stands for ``no trending seasonality", which corresponds to zero coefficients on all seasonal dummy interactions in regression (\ref{evolvereg2}) (in which case it collapses to linear trend and fixed seasonal dummies with serial correlation).  The results  are striking.  There is no evidence for evolving seasonality in AVG; the median AVG $p(nts)$ is 0.73. In contrast, there is strong evidence of  evolving seasonality in DTR; the median DTR $p(nts)$  is 0.03.

\subsection{Serial Correlation}

Estimated AVG and DTR serial correlation coefficients appear in column 6 of Tables \ref{vvv} and \ref{vvv2}, respectively.\footnote{One could also allow for the possibility of more sophisticated forms of serial correlation. There is, for example, some borderline evidence of long memory in the AVG equation (4), as discussed in  \cite{CD2005}. They did not, however, study DTR, and to the best of our knowledge 	long memory has not been explored  in the DTR equation (4).  Some preliminary exploration  allowing for fractionally-integrated ARFIMA(p,d,q)	disturbances produced mixed results, but overall there was some evidence of long memory $(0 {<} d {<} 0.5)$. We leave a full exploration to future work, particularly as there are many issues and	nuances regarding long memory and its relationship to structural change, as emphasized by \cite{DieboldInoue2001}.} All are positive and significant at the one percent level. Their magnitudes, however, are very different.  All those for AVG are around 0.75, whereas all those for DTR are around 0.35.

It is interesting to note that, although the signals in both AVG and DTR are clearly driven by trend, seasonal, and cyclical components, the DTR signal is burried in much more noise.  As shown in column 7   of Tables \ref{vvv} and \ref{vvv2}, respectively, all AVG regression $R^2$ values are around 0.9, whereas all those for DTR are around 0.2.

\section{Concluding Remarks}
\label{concl} 

Climate change is one of the most consequential and pressing issues of our time. We have focused on DTR as an important summary statistic for characterizing climate change, and we have estimated  new stochastic representations of DTR that can capture  its changing seasonality over time and space. Throughout we have also provided parallel contrasting  results for AVG. Many extensions are possible, such as full bivariate modeling of AVG and DTR, or MAX and MIN, and decomposing the shift in the DTR seasonal pattern into underlying shifts in the MAX and MIN patterns. 

More generally, our results may prove useful for assessing and improving structural climate models. Previous research  shows that DTR is a useful metric to help assess the accuracy and degree of fit of global climate models \citep{Braganzaetal2004,Zhouetal2010,lewis2013,Rader2018}. They generally found that these models persistently underestimated the trend in DTR, which was likely related to deficiencies in modeling water vapor and cloud cover processes.  Our new results on the evolving seasonality of DTR may provide an additional, more refined, benchmark for such evaluations.

Our results may also prove useful for assessing financial market efficiency, that is, for assessing whether the temperature forecasts embedded in financial asset prices accurately reflect temperature's underlying dynamics.  It may be of interest to extend existing work based on AVG \citep{NBERw25554} to incorporate our more complete model of AVG dynamics, or to consider multivariate modeling of AVG and DTR, extending the univariate approach undertaken in this paper.

\clearpage
\bibliographystyle{Diebold}
\addcontentsline{toc}{section}{References}
\bibliography{Bibliography}

\clearpage

\appendix
\appendixpage
\addappheadtotoc
\newcounter{saveeqn}
\setcounter{saveeqn}{\value{section}}
\renewcommand{\theequation}{\mbox{\Alph{saveeqn}.\arabic{equation}}} \setcounter{saveeqn}{1}
\setcounter{equation}{0}

\section{Sequential and Joint Regression Results for Philadelphia} \label{regresphl}

\renewcommand{\thefigure}{A\arabic{figure}}

\setcounter{figure}{0}

\begin{figure}[h]
	\caption{PHL Trend Regression, AVG}
	\begin{center}
		\includegraphics[scale=1,trim={0cm 0 0 0},clip]{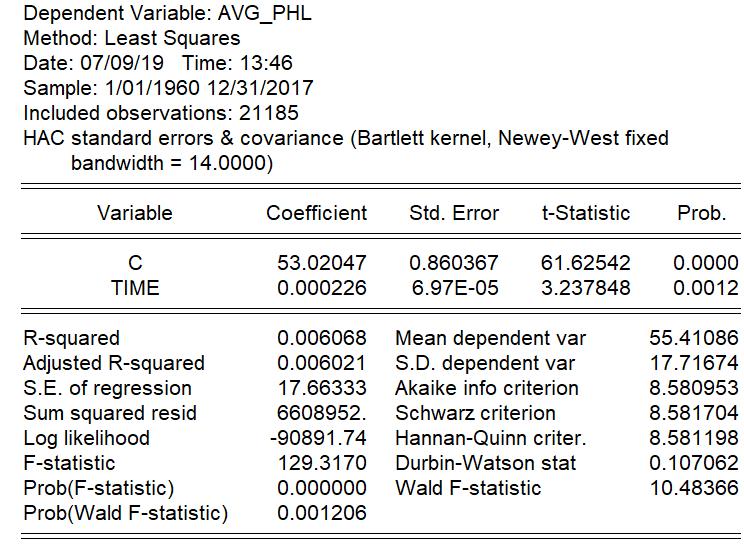}
		\label{tab1_phl}
	\end{center}
\end{figure}

\clearpage

\begin{figure}[h]
	\caption{PHL  Trend Regression, DTR}
	\begin{center}
		\includegraphics[scale=1,trim={0cm 0 0 0},clip]{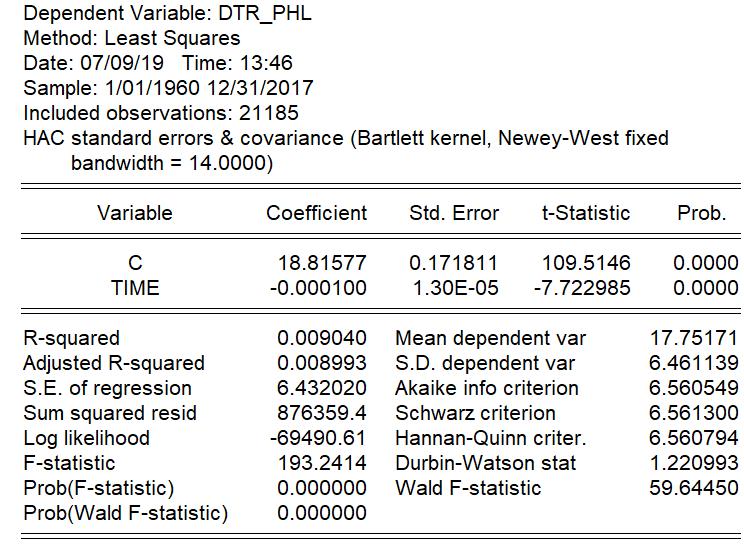}
		\label{tab2_phl}
	\end{center}
\end{figure}

\clearpage

\begin{figure}[h]
	\caption{PHL  Fixed Seasonal Regression, AVG}
	\begin{center}
		\includegraphics[scale=1,trim={0cm 0 0 0},clip]{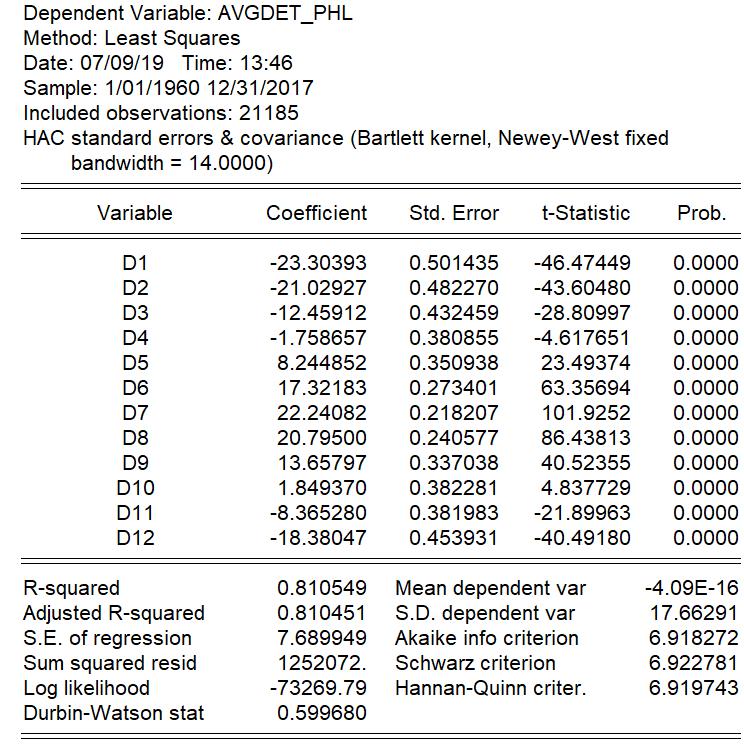}
		\label{tab3_phl}
	
	\footnotesize{Notes: The regression is based on de-trended data. See text for details.}
\end{center}
	
\end{figure}

\clearpage

\begin{figure}[h]
	\caption{PHL  Fixed Seasonal Regression, DTR}
	\begin{center}
		\includegraphics[scale=1,trim={0cm 0 0 0},clip]{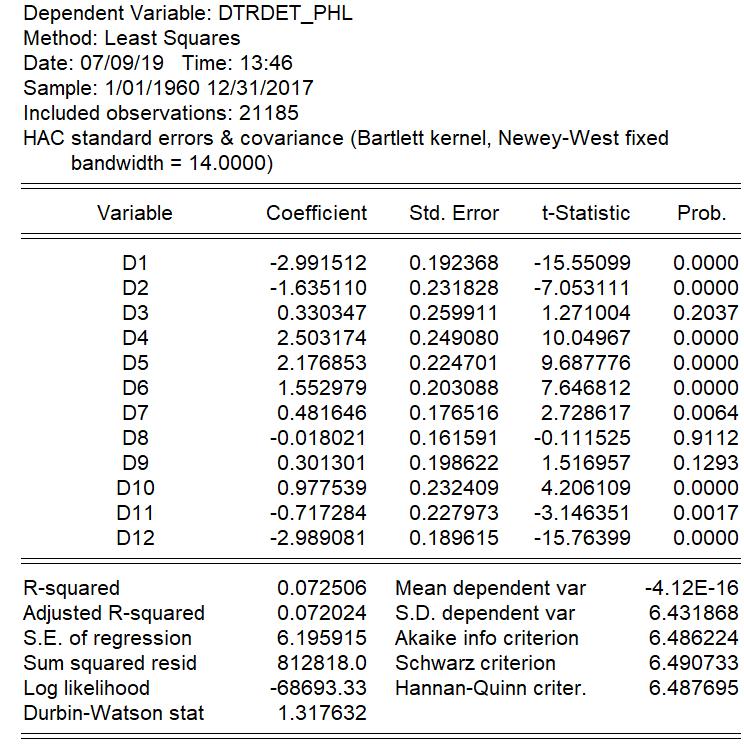}
		\label{tab4_phl}

	\footnotesize{The regression is based on de-trended data. See text for details.}

	\end{center}
\end{figure}

\clearpage

\begin{figure}[h]
	\caption{PHL  Evolving Seasonal Regression, AVG}
	\begin{center}
		\includegraphics[scale=1,trim={0cm 0 0 0},clip]{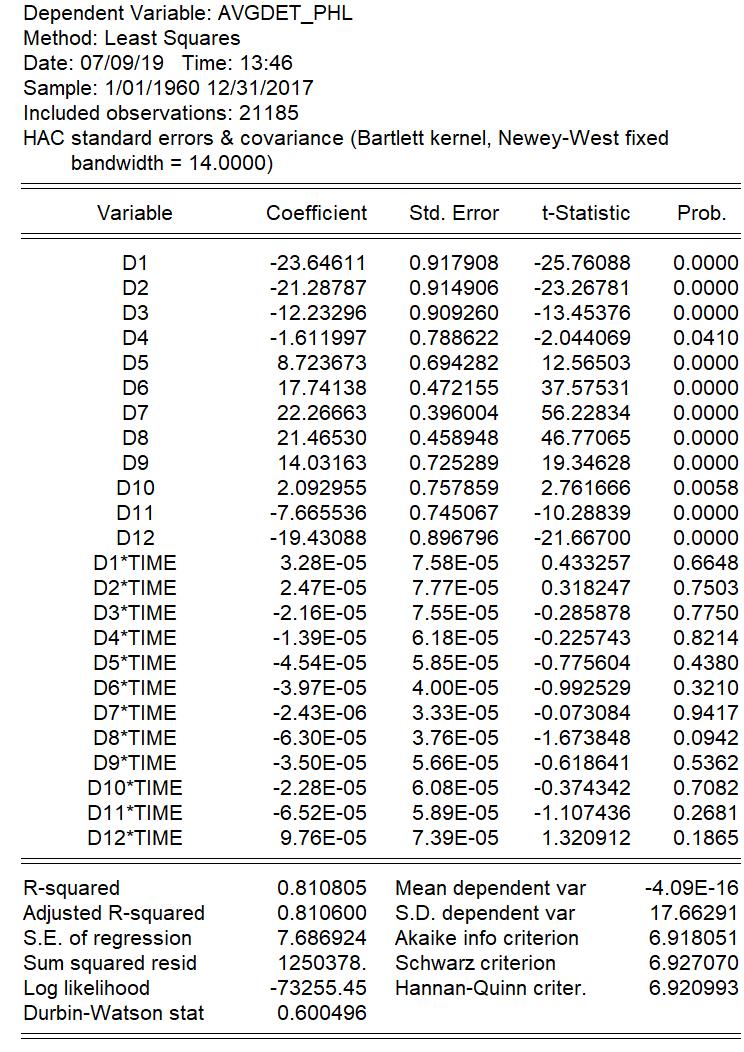}
		\label{tab5_phl}
	
\footnotesize{The regression is based on de-trended data. See text for details.}

\end{center}
	
\end{figure}

\clearpage

\begin{figure}[h]
	\caption{PHL  Evolving Seasonal Regression, DTR}
	\begin{center}
		\includegraphics[scale=1,trim={0cm 0 0 0},clip]{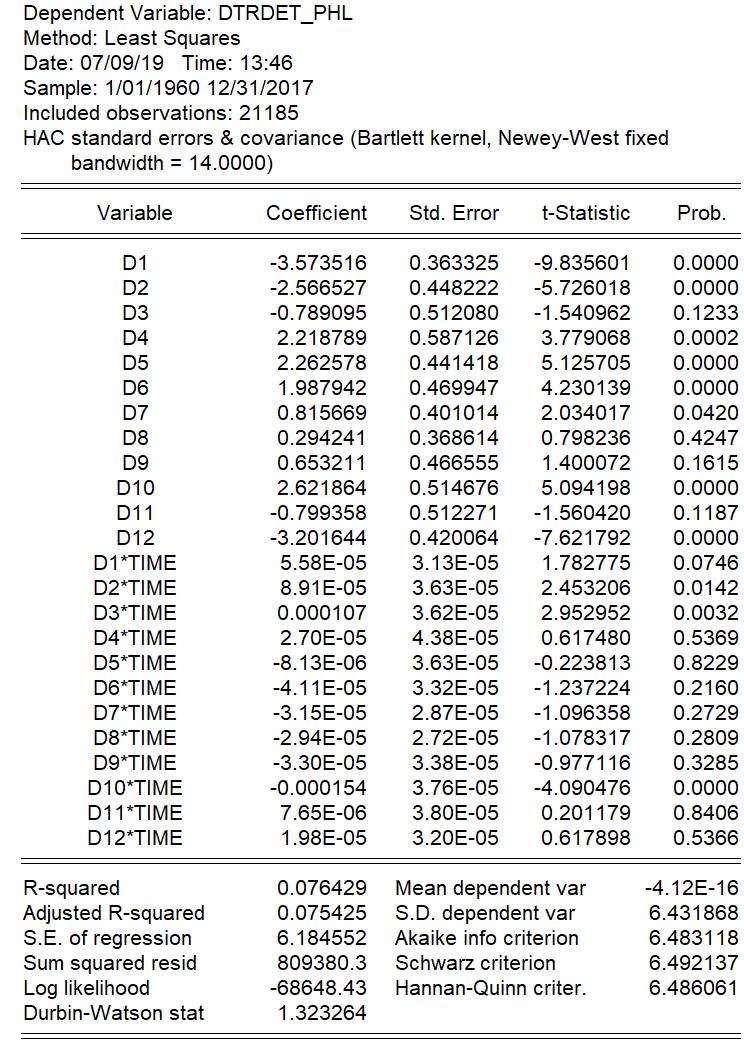}
		\label{tab6_phl}
	
\footnotesize{The regression is based on de-trended data. See text for details.}

\end{center}
	
\end{figure}

\clearpage

\begin{figure}[h]
	\caption{PHL Joint Conditional Mean Regression, AVG}
	\begin{center}
		\includegraphics[scale=1,trim={0cm 0 0 0},clip]{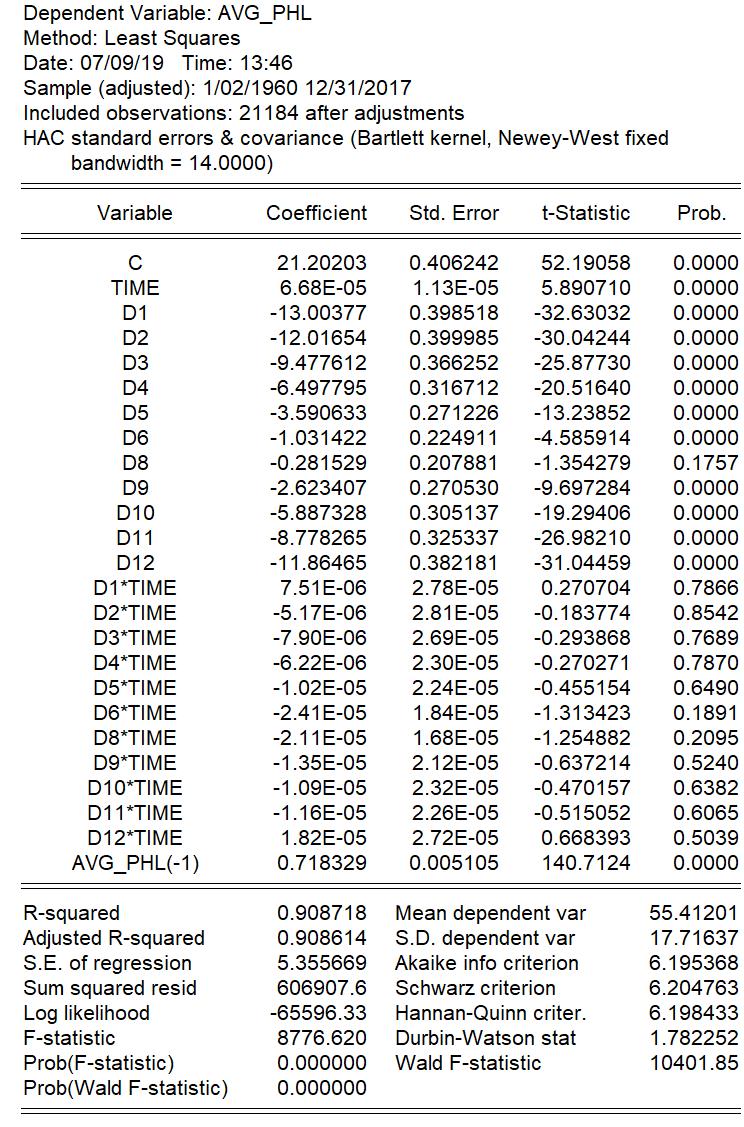}
		\label{tab7_phl}
	\end{center}
\end{figure}

\clearpage

\begin{figure}[h]
	\caption{PHL Joint Conditional Mean Regression, DTR}
	\begin{center}
		\includegraphics[scale=1,trim={0cm 0 0 0},clip]{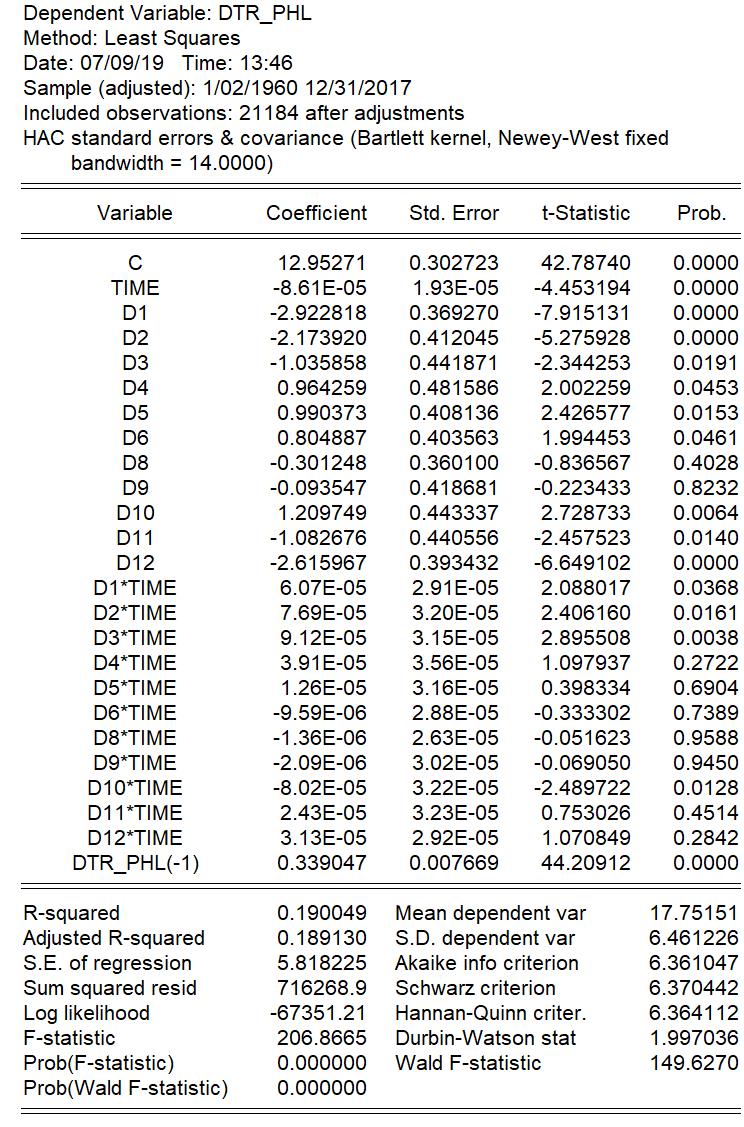}
		\label{tab8_phl}
	\end{center}
\end{figure}

\clearpage

%
%
%
%
%
%

\end{document}